\documentclass[referee]{raa}            

\usepackage{graphicx,times}             
\input{epsf.sty}

\begin{document}

\def\lsim{\lower.5ex\hbox{$\; \buildrel < \over \sim \;$}}
\def\gsim{\lower.5ex\hbox{$\; \buildrel > \over \sim \;$}}

\title{2.5-dimensional solution of the advective accretion disk: A self-similar approach}

\setcounter{page}{1}

\author{ Shubhrangshu Ghosh\inst{1} \and Banibrata Mukhopadhyay\inst{2}}
\institute{ Indian Institute of Astrophysics, Koramangala, Bangalore 560034, India;
{\it sghosh@iiap.res.in}\\ \and
Astronomy and Astrophysics Programme, Department of Physics, Indian Institute of Science,
Bangalore 560012, India; {\it bm@physics.iisc.ernet.in}
}

\def\ch{\lower-0.55ex\hbox{--}\kern-0.55em{\lower0.15ex\hbox{$h$}}}
\def\lh{\lower-0.55ex\hbox{--}\kern-0.55em{\lower0.15ex\hbox{$\lambda$}}}       
\def\n{\nonumber}

\abstract{
We provide a 2.5-dimensional solution to a complete set of 
viscous hydrodynamical equations describing accretion-induced outflow and then plausible jet around black holes/compact 
objects. We prescribe a self-consistent advective disk-outflow
coupling model, which explicitly includes the information of vertical 
flux. Inter-connecting dynamics of inflow-outflow system 
essentially upholds the conservation laws. We provide a 
set of analytical family of solutions through the self-similar approach. The flow parameters of  
the disk-outflow system 
depend strongly on viscosity parameter $\alpha$ and cooling 
factor $f$. 
}

\keywords{ accretion, accretion disk, black hole physics, hydrodynamics,
galaxies: jets}
\maketitle


\section{Introduction}

Most extragalactic radio sources are expected to form 
around spinning massive black holes (Meier et al. \cite{msu}, Meier \cite{md}). The immense amount of matter, forming an accretion disk, is being accreted 
either from the interstellar medium or from its companion star. In these systems, the relativistic 
outflowing matter should come
 only from the inner regions of the accretion disk unlike stellar outflows.
This is particularly suggestive for quasars or the micro-quasars which do not 
have an atmosphere of their own. 
Fender, Belloni \& Gallo (\cite{fb}) suggested a semi-quantitative model for the jet in black hole X-ray binaries 
where a correlation between the radio and the X-ray emission was estimated. 
Vadawale et al. (\cite{vs}) established  the X-ray and radio properties of micro-quasar GRS 1915+105. 
Time 
dependent interaction between the jet and the inner disk (e.g. Ueda et al. \cite{uy}) was evident from the observations of simultaneous 
X-ray/IR flares from a black hole/relativistic system. Rawlings \& Saunders (\cite{rs}) found a 
strong correlation between the narrow-line and radio luminosity in FRII type radio galaxies. This implies 
that the production of optical line emission and  
large-scale radio emission are intrinsically linked. 
Therefore, the outflows or jets are expected to correlate with the 
disk controlling 
the accretion process, precisely the accretion dynamics around a central star. The jets or outflows extract 
matter, energy and angular momentum from the disk. 

Thus it is now clear that these two apparently dissimilar objects are 
related each other. In principle, one should study the disks and outflows leading to jets 
in a unified manner, which cannot be dealt as separate flow dynamics. 
However, there are few models which simultaneously study the accretion-outflow 
dynamics on the same platform. Chakrabarti \& Bhaskaran (\cite{cb}) attempted to 
correlate the collimated bipolar outflows with the disk through a 
simplified model based on the ambipolar diffusion approximation and self-similarity 
in the radial direction. Blandford \& Begelman (\cite{bb1}) modified the 
ADAF solution, originally proposed by Narayan \& Yi (\cite{ny94}) 
where the accretion flow is well below the Eddington limit, 
by including an outflow/wind which carries 
mass, angular momentum and energy from the accretion disk. They later extended their 
work to two-dimensional adiabatic flow (Blandford \& Begelman \cite{bb2}). 
Although a new branch of wind solutions was discovered, that does not 
include the vertical fluxes in the hydrodynamical equations. 
The ADAF model (Narayan \& Yi \cite{ny94,ny95})
explained the under-luminous accreting sources. The interesting aspect of 
the ADAF model is that the Bernoulli's parameter at all radii (within the acceptable location 
of the validity of the self-similar approach) is positive, which leads to conceive that
the outflows and jets might emanate from the advective disk. 
Later on, stability of the solution under perturbation was studied with 
the inclusion of Coriolis force by Prasanna \& Mukhopadhyay (\cite{pm}). In recent times a few simulations 
on disk-outflow coupling have been cultivated (Nishikawa et al. \cite{NI}, McKinney \& Narayan \cite{MN}). 
However, the results are strongly dependent on the initial conditions (Ustyugova et al. \cite{U}) and it is difficult to simultaneously simulate the disk and the outflow regions because the time scales 
of the accretion and outflow are in general very different. Moreover, in these simulations how 
the matter gets deflected from the equatorial plane has been studied 
largely in the Keplerian regime. 

In recent years, there have been a discovery of unusual class of compact sources, the ultra-luminous 
X-ray sources (ULX), in the nearby 
star forming galaxies (Katz \cite{kt}, Fabbiano et al. \cite{fb2},
Kaaret et al. \cite{kp}, Colbert \& Ptak \cite{cp}, Miller et al. \cite{mf},
Begelman et al. \cite{bk}). These are optically 
thick, radiation pressure dominated 
systems with strong advection and the matter is strongly ejected out from the disk 
in the form of outflows/jets by strong radiation pressure. 
Using a slim disk model, Abramowicz et al. (\cite{ac}) discovered a new
branch of solution at a super-critical rate which is stable and optically thick. 
A model for super-critical accretion with 
advection was attempted by Lipunova (\cite{lg}). Ohsuga et al. (\cite{om}) have 
emphasized the importance of advective flows in the super-Eddington, radiation 
pressure dominated disk with photon trapping. 
Hence, these two opposite paradigms of black hole activities reveal a 
profound inter-connection between the inflow parameters and the
outflows leading to jets, especially in the advective regime, 
which the standard optically thick Keplerian disk theory fails to explain.  

In the present work, without assuming a geometrically
thin disk structure, we prescribe a new model for the accretion-induced 
outflow leading to jet. We construct the inflow-outflow correlation model in a more self-consistent manner. 
The contribution of magnetic field is neglected at the first instant. The magnetic field is more 
important to explain the 
collimation and acceleration of jet (apart from ultra luminous sources)\footnote{The 
conservation equations should remain valid amidst of the nature of model.}. 
The present model can, not only extend our model from quasars to micro-quasars, but also to 
neutron star X-ray binaries and in general to many 
sources with outflows from the disk. However, to describe the flow dynamics and 
consequently outflows in protostellar objects the standard Keplerian 
disk model itself is enough.
Our unification scheme is based on the fact that the astrophysical outflow and jet, its 
underlying disk and its inter-related dynamics at all scales, obey same physical laws. 

We arrange our paper in the following manner. In the next 
section, we formulate our model equations for the 
accretion-induced outflow. In \S 3, we present a complete 
analytical, but self-similar, solution of our model. Next, we study 
the properties of the class of solution in \S 4. In \S 5 we end with a 
discussion and summary. 

\section{Disk-outflow correlation and model equations}

We assume the disk to be 
steady and axisymmetric. For a generalized geometrically thick advective disk, we consider the 
$r\phi-$, $\phi z-$ and $rz-$ components of the shearing stress. The remaining stresses are believed to be 
negligible which do not significantly contribute to 
control the disk-outflow
dynamics. The flow parameters $v_r, \lambda, v_z$, $c_s$, 
$\rho$ and $P$ are considered to be functions 
of both radial and vertical coordinates, which are
radial velocity, specific 
angular momentum, 
vertical velocity, adiabatic sound speed, mass density and pressure 
respectively. Here, 
throughout our calculations, we express radial and vertical 
coordinate in the unit of $2GM/c^2$, where $M$ is mass of the central 
star, $G$ is the gravitational constant and $c$ is speed of light. 
We also express velocities in the unit of speed of light and specific angular 
momentum in $2GM/c$. The mass of the disk is assumed to be much 
less than that of the central object, hence the disk is not self-gravitating.
Therefore, the general disk-outflow coupled 
equations are given below.

(a) Mass transfer:

\begin{eqnarray}
\frac{1}{r} \frac{\partial}{\partial r} (r \rho v_r) \,+ \, \frac{\partial}{\partial z} (\rho v_z) =  0.
\label{1a}
\end{eqnarray}

(b) Radial momentum balance:


\begin{eqnarray}
v_r \frac{\partial v_r}{\partial r} \,+ \, v_z \frac{\partial v_r}{\partial z} \, - \, \frac{\lambda^{2}}{r^3} \, + \, F_{Gr} \, + \, \frac{1}{\rho} \frac{\partial P}{\partial r} \, - \, \frac{1}{\rho}\frac{\partial W_{rz}}{\partial z} \, = \, 0,
\label{2a}
\end{eqnarray}
where $W_{rz}$ is the $rz^{th}$ component of the stress 
tensor, when we consider the shear stress tensor is symmetric (Landau \& Lifshitz \cite{ll}) 
and $F_{Gr}$ is radial component of the gravitational force. 
To understand the importance of the 
term ${\partial W_{rz}}/{\partial z}$ in the above equation, we 
compare it with ${\partial P}/{\partial r}$ as
\begin{eqnarray}
\bigg|\frac{\partial W_{rz}}{\partial z}\bigg|\bigg/ \bigg|\frac{\partial P}{\partial r}\bigg| \, \sim \,\frac{r}{h} \frac{\nu_t}{c^{2}_{s}} \bigg|\biggl(\frac{v_r}{h} \, + \, \frac{v_z}{r}\biggr) \bigg|,
\label{2b}
\end{eqnarray}
where we use a generic order of magnitude relation 
${\partial A}/{\partial x_j} \approx O (A/x_j)$; $A$ denotes 
any independent quantity as a function of an arbitrary coordinate variable 
$x_j$, $h(r)$  is the disk half-thickness. Note that we do not identify 
$h$ here as a hydrostatic 
scale height, instead the photospheric 
height where the disk is coupled to the 
corona, $\nu_t$ is the turbulent kinematic viscosity. 
With $c^{2}_{s} \sim P/\rho$ and from eqn. (\ref{1a}) we obtain
\begin{eqnarray}
\big|\frac{v_z}{v_r} \big| \, \sim \, \frac{h}{r}.
\label{2c}
\end{eqnarray}
However, we can write from eqn. (\ref{3b}) (as described 
below)
\begin{eqnarray}
v_r \,  \sim \, \frac{\nu_t}{r}.
\label{2d}
\end{eqnarray}
Using eqns. (\ref{2c}) and (\ref{2d}), and assuming 
an isotropic distribution of turbulence\footnote{In reality 
turbulence is generally anisotropic for a thick disk.} 
such that $\nu_t \sim \alpha c_s h$, where $\alpha$ $(\alpha \leq 1)$ is the 
Shakura-Sunyaev 
viscosity parameter (Shakura \& Sunyaev \cite{ss73}),  
eqn. (\ref{2b}) reduces to 
\begin{eqnarray}
\bigg|\frac{\partial W_{rz}}{\partial z}\bigg|\bigg/\bigg|\frac{\partial P}{\partial r}\bigg| \, \sim \, \alpha^{2} \, + \, \alpha^{2} \left(\frac{h}{r}\right)^2.
\label{2e}
\end{eqnarray}
For a reasonable value of $h \sim r/2$ and $\alpha$, the second 
term on the right hand side of eqn. (\ref{2e}) can be neglected. Thus we 
retain with $({\partial W_{rz}}/{\partial z})\bigg/\frac{\partial P}{\partial r} \sim \alpha^{2}$; ${\partial W_{rz}}/{\partial z}$ can not be neglected.  
In order to determine $W_{rz}$, we derive a 
simplified relation of $W_{rz}$ with $W_{r\phi}$, which is the $r \phi^{th}$ component of the stress tensor, from the order of magnitude 
analysis and obtain 
$W_{rz} \sim \alpha W_{r \phi}\,h/r$

As the disk has a significant radial flow, we include ram pressure along with 
gas pressure (Mukhopadhyay \& Ghosh \cite{mb}) in the equations and write $W_{r\phi} = -\alpha (P+\rho v^{2}_{r})$. The radial momentum equation of the disk-induced outflow/jet thus reduces to 
\begin{eqnarray}
v_r \frac{\partial v_r}{\partial r} \, + \, v_z \frac{\partial v_r}{\partial z} \, - \, \frac{\lambda^{2}}{r^3} \, + \, F_{Gr} \, + \, \frac{1}{\rho} \frac{\partial P}{\partial r} \, + \, \frac{\alpha^2}{\rho} \frac{\partial}{\partial z} [\frac{z}{r} (P+\rho v^{2}_{r})] \, = \, 0, 
\label{2f}
\end{eqnarray}
where we have used the fact that for a thick disk, in general, $h \sim z$.

(c) Azimuthal momentum balance:

\begin{eqnarray}
v_r \frac{\partial \lambda}{\partial r} \, + \, v_z \frac{\partial \lambda}{\partial z} \, = \, \frac{1}{\rho r} \frac{\partial}{\partial r} (r^2 W_{r\phi}) \, + \, \frac{r}{\rho} \frac{\partial W_{\phi z}}{\partial z}.
\label{3a}
\end{eqnarray}
The first term on the right hand side signifies the outward transport 
of angular momentum in the radial direction and the second term 
in the vertical direction due to the turbulent stress. 

If $W_{r\phi}$ dominants the angular momentum transport, 
with the use of mass conservation eqn. (\ref{1a}), we obtain
\begin{eqnarray}
|v_r| \sim \frac{|W_{r \phi}|}{\rho v_{\phi}}.
\label{3b}
\end{eqnarray}
If, on the other hand, $W_{\phi z}$ dominates the angular momentum transport, 
then we obtain 
\begin{eqnarray}
|v_z| \sim \frac{|W_{\phi z}|}{\rho v_{\phi}}.
\label{3c}
\end{eqnarray} 
Now comparing eqns. (\ref{3b}) and (\ref{3c}) and with the use of 
mass conservation eqn. (\ref{1a}), we can write 
\begin{eqnarray}
W_{\phi z} \, \sim \, \frac{h}{r} \, W_{r \phi}.
\label{3d}
\end{eqnarray}
Therefore, the azimuthal equation 
reduces to  
\begin{eqnarray}
v_r \frac{\partial \lambda}{\partial r} \, + \, v_z \frac{\partial \lambda}{\partial z} \, + \, \frac{\alpha}{r \rho}\Biggl[\frac{\partial}{\partial r} [r^2 (P + \rho v^{2}_{r})] \, + \, r^2 \frac{\partial}{\partial z} [\frac{z}{r} (P+\rho v^{2}_{r})]\Biggr] \, = \, 0.
\label{3e}
\end{eqnarray}

(d) Vertical momentum balance:

\begin{eqnarray}
v_r \frac{\partial v_z}{\partial r} \, + \, v_z \frac{\partial v_z}{\partial z} \, +  \, F_{Gz} \, + \, \frac{1}{\rho} \frac{\partial P}{\partial z} \, - \, \frac{1}{r \rho} \frac{\partial }{\partial r} (r W_{rz}) \, = \, 0,
\label{4a}
\end{eqnarray}
where $F_{Gz}$ is the vertical component of the gravitational force.
As before we estimate $\bigg|\frac{1}{r} \frac{\partial}{\partial r} (r W_{rz}) \bigg|\bigg/\bigg|\frac{\partial P}{\partial z}\bigg| \sim \alpha^2 (\frac{h^2}{r^2} \, + \, \frac{h^4}{r^4})$. For $h \sim r/2$ and 
reasonable $\alpha$, the 
quantity is negligible. The eqn. (\ref{4a}) thus reduces to 

\begin{eqnarray}
v_r \frac{\partial v_z}{\partial r} \, + \, v_z \frac{\partial v_z}{\partial z}
\,+ \, F_{Gz} \, + \, \frac{1}{\rho} \frac{\partial P}{\partial z} \, = \, 0.
\label{4b}
\end{eqnarray}
In absence of first term, the equation leads to 
a mean vertical outflow from the disk. On the other hand, if there is no outflow and jet: 
$v_z = 0$, then eqn. (\ref{4b}) reduces to the well 
known hydrostatic equilibrium 
condition in the disk, from where one can calculate the 
hydrostatic disk-scale height.   


(e) Energy conservation:

For an accretion-induced outflow the energy budget can be computed by 
\begin{eqnarray}
\frac{1}{r} \frac{\partial}{\partial r} (r {\mathcal F}_r) \, + \, 
\frac{\partial {\mathcal F}_z}{\partial z}=0, 
\label{5a}
\end{eqnarray}
where ${\mathcal F}_r$ and ${\mathcal F}_z$ are the radial and vertical 
components of the total energy flux ${\mathcal F}_i$ given by 
\begin{eqnarray}
{\mathcal F}_i = \rho v_i \biggl(\frac{v^2}{2} + \frac{\gamma}{\gamma-1} \frac{P}{\rho} + \phi_G \biggr) 
- v_j W_{ij} + F_i,
\label{5b}
\end{eqnarray} 
where we neglect the molecular heat conduction 
as this is insignificant in accretion disks 
and the nuclear heat generation/absorption (Mukhopadhyay \& Chakrabarti \cite{mc}) for 
mathematical simplicity.
Here $v^2=v_r^2+\lambda^2/r^2+v_z^2$, $\phi_G$ is the gravitational
potential, $F_i$ the radiative flux from 
the disk surface, $W_{ij}$ the generalized stress tensor, $\gamma$ 
the gas constant:  $4/3 < \gamma < 5/3$.

Using eqns. (\ref{1a}), (\ref{2a}), (\ref{3a}), 
(\ref{4a}) $\&$ (\ref{5b}), we obtain the disk-energy 
equation from (\ref{5a}) as
\begin{eqnarray}
\rho v \cdot \nabla s \, = \, \frac{v_r}{\Gamma_3-1} \biggl[\frac{\partial P}{\partial r} - \Gamma_1 \frac{P}{\rho} \frac{\partial \rho}{\partial r}\biggr] \, + \, \frac{v_z}{\Gamma_3-1} \biggl[\frac{\partial P}{\partial z} - \Gamma_1 \frac{P}{\rho} \frac{\partial \rho}{\partial z}\biggr] \, = \, Q^{+} \, - \, Q^{-} \, = \, f Q^{+} , 
\label{5c}
\end{eqnarray}
where $s$ is entropy density. Here we assume the energy released $Q^{-}$ due to radiative loss from the disk is proportional 
to the viscous heat generated, $Q^{+}$, where $f$ is the cooling factor 
incorporating any kind of outflow and jet which is 
close to $0$ and $1$ for the flow with efficient and 
inefficient cooling respectively. The first and the second terms on the left hand side of the above 
equation represent the radial and vertical advection of the flow respectively, 
where we define (Cox \& Giuli \cite{cg}, Mukhopadhyay \& Ghosh \cite{mb})
\begin{eqnarray}
\nonumber
\Gamma_3=1+\frac{\Gamma_1-\beta}{4-3\beta},\\
\nonumber
\Gamma_1=\beta+\frac{(4-3\beta)^2(\gamma-1)}{\beta+12(\gamma-1)(1-\beta)},\\
\beta=\frac{\rho k_B T/\mu m_p}{\bar{a} T^4/3+\rho k_B T/\mu m_p},
\label{5d}
\end{eqnarray}
with $\beta = \frac{6 \gamma -8}{3 \gamma -3}$, the ratio of gas 
pressure to total pressure which is close to $0$ for extreme radiation 
dominated flow $(\gamma = 4/3)$ and to $1$ for extreme 
gas dominated flow $(\gamma = 5/3)$, $\bar{a}$ is Stefan constant, $m_p$ 
is mass of the proton, $T$ is proton temperature, $k_B$ is the Boltzmann 
constant, $\mu$ is average molecular weight. 

In eqn. (\ref{5c}), $Q^{+}= {W^{2}_{ij}}/{\eta_t}$, $\eta_t$ is the 
coefficient of turbulent viscosity. Thus for our case 
\begin{eqnarray}
Q^{+} \, = \, \frac{1}{\eta_t} (W^{2}_{r\phi} + W^{2}_{\phi z} + W^{2}_{rz}).
\label{5e}
\end{eqnarray}
As before, it can easily be shown that the contribution of $W_{rz}$ is much less 
than that due to ${W_{r\phi}}$ \bigg(${W^{2}_{rz}}/{W^{2}_{r\phi}} 
\sim \alpha^2 [\frac{h^2}{r^2} + 
\frac{h^4}{r^4}]$ \bigg). $W_{\phi z}$ contributes to the 
additional viscous heating in a geometrically thick 
advective disk with vertical outflow. Using mixed 
shear stress formalism (Chakrabarti \cite{cs}) and 
approximating $W_{\phi z}$ in terms of $W_{r \phi}$ as given by eqn. (\ref{3d}), eqn. (\ref{5c}) reduces to 
\begin{eqnarray}
\frac{v_r}{\Gamma_3-1} \biggl[\frac{\partial P}{\partial r} - \Gamma_1 \frac{P}{\rho} \frac{\partial \rho}{\partial r}\biggr] \, + \, \frac{v_z}{\Gamma_3-1} \biggl[\frac{\partial P}{\partial z} - \Gamma_1 \frac{P}{\rho} \frac{\partial \rho}{\partial z}\biggr] \, = \, - f \alpha (P+\rho v^{2}_{r}) \frac{1}{r} \biggl(\frac{\partial \lambda}{\partial r} -2 \frac{\lambda}{r} + \frac{z}{r} 
\frac{\partial \lambda}{\partial z}\biggr).
\label{5e}
\end{eqnarray}










\section{Solution and self-similarity}

We follow the self-similar approach to solve the
equations in obtaining the class of solutions. For
the present purpose we seek for a generalized self-similar 
solution, unlike the previous case (Narayan \& Yi \cite{ny94}), where variation
of the flow parameters as functions of vertical coordinate
along with radial coordinate has been invoked
for a coupled set of disk-outflow equations of the form 
\begin{eqnarray}
v_r (r,z) \, = \, v_{r0} r^{u} z^{a}, \, \, \lambda (r, z) \, = \, 
\lambda_0 r^{v} z^{b}, \, \, v_z (r, z) \, = \, v_{z0} r^{w} z^{d}, \, \, c_s (r, z) = c_{s0} r^{g} z^{s},
\label{70a}
\end{eqnarray} 
where $v_{r0}$, $\lambda_0$, $v_{z0}$ and $c_{s0}$ are the dimensionless 
coefficients which will be evaluated from the conservation 
equations. We determine the exponents $u,a,v,b,w,d,g,s$ by self comparison 
of various terms in the equations. 
Assuming the flow to be polytropic, as most likely the disk is,
we consider the adiabatic equation 
of state as $P=k \rho^{\gamma}$, where $\gamma = 1+1/n$, 
$n$ is the polytropic index of the flow, while 
the adiabatic sound speed $c_{s} = \sqrt{\gamma\,P/\rho}$.

We propose a generalized gravitational potential 
$\phi_G (r,z)=-(r^{-1}-\frac{1}{k+2}r^{-3} z^2)z^k$, where the index
$k$ induces the variation along z-axis which we determine self-consistently. 
When the disk does not have strong outflow and jet $k=0$, 
and $\phi_G$ reduces to conventional Newtonian potential upto the second order 
in $(z/r)$. 

Substituting the solutions 
from eqn. (\ref{70a}) in eqns. (\ref{1a}) and (\ref{2f}) and
comparing the exponents of $r$ and $z$ we 
obtain $w \, = \, u-1, a \, = \, s, \, d \, = \, a+1, \,
u \, = \, -1/2, \, v \, = \, 1/2, \, g \, = \, -1/2, b \, = \, a, \, 
k \, = \, 2a$. 

Equations (\ref{1a}), (\ref{2f}), (\ref{3e}) and (\ref{5e}) can now 
be written, with the use of eqn. (\ref{70a}), respectively

\begin{eqnarray}
v_{r0} + 2 \biggl(\frac{2 a n + a +1}{1-2n}\biggr) v_{zo} \, = \, 0,
\label{70b}
\end{eqnarray}

\begin{eqnarray}
\biggl[\frac{1}{2} - \gamma \alpha^2 [2 a (n+1) + 1]\biggr] v^{2}_{r0}  +  \biggl[n- \alpha^2 [2 a (n+1) + 1]\biggr] c^2_{s0} + \lambda^{2}_{0} - a v_{r0} v_{z0}  - 1  =  0,
\label{70c}
\end{eqnarray}

\begin{eqnarray}
\biggl(\frac{1}{2} v_{r0} + a v_{z0}\biggr) \lambda_0 + \alpha \biggl[(n+1)(2a-1) +3\biggr]\biggl(v^{2}_{r0} + \frac{n}{n+1} c^{2}_{s0} \biggr)=0
\label{70d}
\end{eqnarray}

and

\begin{eqnarray}
\frac{n(\Gamma_1 -1) - 1}{\Gamma_3-1} \biggl[\frac{1}{2} v_{r0} - a v_{z0}\biggr] c^{2}_{s0} \, + \, \frac{1}{2} f \alpha \biggl(a-\frac{3}{2}\biggr) \biggl[\frac{n+1}{n} v^{2}_{r0} + c^{2}_{s0}\biggr] \lambda_0=0.
\label{70e}
\end{eqnarray}
Solving eqns. (\ref{70b})-(\ref{70e}) 
we compute the coefficients of eqn. (\ref{70a}) 
\begin{eqnarray}
v_{r0} \, = \, \frac{\mathcal D}{\Biggl[{\mathcal B} {\mathcal X} + {\mathcal D} ({\mathcal A} {\mathcal D} - a) +  {\mathcal K}^{2} \frac{{\mathcal X}^{2}}{{\mathcal G}^{2}}\Biggr]^{1/2}},
\label{70g}
\end{eqnarray}
\begin{eqnarray}
\lambda_{0}\, = \, \frac{{\mathcal H} {\mathcal X}}{{\mathcal G} \Biggl[{\mathcal B} {\mathcal X} + {\mathcal D} ({\mathcal A} {\mathcal D} - a) +  {\mathcal K}^{2} \frac{{\mathcal X}^{2}}{{\mathcal G}^{2}}\Biggr]^{1/2}},
\label{70h}
\end{eqnarray}
\begin{eqnarray}
v_{z0}\, = \, \frac{1}{\Biggl[{\mathcal B} {\mathcal X} + {\mathcal D} ({\mathcal A} {\mathcal D} - a) +  {\mathcal K}^{2} \frac{{\mathcal X}^{2}}{{\mathcal G}^{2}}\Biggr]^{1/2}}, 
\label{70i}
\end{eqnarray}
and 
\begin{eqnarray}
c_{s0}\, = \, \frac{{\mathcal X}^{1/2}}{\Biggl[{\mathcal B} {\mathcal X} + {\mathcal D} ({\mathcal A} {\mathcal D} - a) +  {\mathcal K}^{2} \frac{{\mathcal X}^{2}}{{\mathcal G}^{2}}\Biggr]^{1/2}}, 
\label{70k}
\end{eqnarray}
where ${\mathcal G} = {\mathcal X} +\gamma \, {\mathcal D}^{2}$, $\mathcal X$ is 
given by 
\begin{eqnarray}
{\mathcal X} = -\frac{\gamma}{4 \mathcal E} \Biggl[4 {\mathcal D}^2 {\mathcal E} + (2a+{\mathcal D}) {\mathcal K} \, \, \biggl[1+\biggl(1+\frac{8{\mathcal D}^2 {\mathcal E}}{(2a+{\mathcal D}) {\mathcal K}}\biggr)^{1/2}\biggr]\Biggr]
\label{70m}
\end{eqnarray} 
and $\mathcal A$, $\mathcal B$, $\mathcal D$, $\mathcal E$, $\mathcal H$ and 
$\mathcal K$ are 
represented as 
\begin{eqnarray}
\nonumber
{\mathcal A}=\biggl[\frac{1}{2} - \gamma \alpha^{2} [2 a(n+1) +1]\biggr],\,\, 
{\mathcal B}=\biggl[n - \alpha^{2} [2 a(n+1) +1]\biggr],\\ 
\nonumber
{\mathcal D}=2 (2 \, a \, n+a+1)/(2n-1),\,\, {\mathcal E}=\alpha \bigg[(n+1)(2 a -1) +3 \bigg],\\
{\mathcal H} =\biggl[\frac{n (\Gamma_1-1) -1}{\Gamma_3-1}\biggr]
(a-\frac{1}{2} {\mathcal D})/\frac{1}{2} f \alpha (a-\frac{3}{2}),\,\,
{\mathcal K}=(a-{\mathcal D}/2).
\label{cons}
\end{eqnarray}

The generalized Bernoulli equation is then 
\begin{eqnarray}
\biggl[\frac{1}{2} \biggl(v^{2}_{r0} + \lambda^{2}_{0} r^{-2} + v^{2}_{z0} r^{-2} z^{2}\biggr) + n c^{2}_{s0} -\biggl(1-\frac{1}{2(1+a)} r^{-2} z^2 \biggr) \biggr] r^{-1} z^{2a} = B_E, 
\label{70q}
\end{eqnarray}
where $B_E$ is the Bernoulli constant. 

The above solutions can explain both super-critical and sub-critical 
accretion flows, where the flow is more likely to be strongly advective 
with strong possibility of the outflow and jet. 
Super-critical accretion, of the order of
$\dot M \gsim (10^{-3}- 10^{-6}) {M_{\odot}}/{yr}$, corresponds
to high luminosity sources with mass of the central star $M \sim 10 M_{\odot}$.
In this case, the flow is expected to be radiation pressure dominated 
with maximum physically plausible $\gamma$ is $1.444$, 
corresponding to $P_r \sim P_g$.

To determine the exponent $a$, we vertically integrate eqn. (\ref{4b}) from 
$-h$ to $+h$ after substituting the solutions given by eqns. (\ref{70a}),
(\ref{70g})-(\ref{70k}). 
As the outflow is not likely to emanate from the equatorial plane,
the solution looses its relevance there because the torque due to $W_{\phi z}$ 
exerted on the matter is zero.
However, from a certain finite height $h_0$,
they are relevant describing a disk-outflow system.
We consider $h/r \sim t \leq 1$ and $h_{0}/r \sim t_0 \ll 1$, 
where $t$ and $t_0$ are kept 
constant throughout our analysis. The realistic flow, when
$v_r<0$ and $v_z>0$, demands that $a$ cannot be positive. 
This helps us to fix the boundary condition
of the outflow in the vertical direction. We demand a situation for 
which $t_0$ is least to yield 
a physically realistic $a$. For super-critical flows exhibiting
high luminosity sources, using eqn. (\ref{70b}) 
we obtain a most physically acceptable solution 
for $a$ given by $a \sim -(2/11 + \epsilon)$ for an appropriate 
$t_0  \sim 0.02$ corresponding to a reasonable $t \sim 0.5$, when
$\epsilon$ is a very small number $ \lsim 10^{-3}$. For sub-critical flows
exhibiting under-luminous sources, on the other hand,
the highly sub-critical mass accretion rate 
$\dot M \leq (10^{-10}-10^{-12}) {M_{\odot}}/{yr}$ or 
$\dot M \leq (10^{-5}-10^{-7}) {M_{\odot}}/{yr}$ corresponding to
black holes of mass $M \sim 10 M_{\odot}$ or $M \sim 10^6 M_{\odot}$ 
respectively, for which $P_g \gg P_r$ and $\gamma \lsim 5/3$. 
With a similar argument as above we obtain here 
$a \sim -(1/4 + \epsilon)$  for an appropriate $t_0 \sim 0.07$ corresponding 
to $t \sim 0.5$.

\section{Properties of self-similar solutions}

Here we study the properties
of our solution for cases of under-luminous and
highly luminous sources at sub-critical and super-critical accretion
rates respectively. The standard model of Shakura \& Sunyaev (\cite{ss73}) 
is ineffective  
to describe these two cases of the geometrically thick advective disks having a substantial 
outflow. We describe a typical set of solutions for the accretion-induced outflow using 
the flow parameters 
 obtained in the last section in these two opposite paradigms. A detailed family of
solutions and their observational implications will be discussed elsewhere (Ghosh et al. \cite{4au}). 

\subsection {Super-critical accretion regime}

Let us consider a case where radiation pressure $P_r$ dominates 
over gas pressure $P_g$ for the flow with
high Eddington-accretion rate. We choose $\gamma \sim 1.4$ 
corresponding to $\beta \sim 1/3$, appropriate for the above class of flow. 
The flow is radiation trapped, optically thick and hot. Figure 1 describes 
the variations 
of flow parameters as functions of radial and vertical coordinates for two
values of $f$ at a typical $\alpha$. We see that $v_r$, $v_z$, $\lambda$ and 
$c_s$ fall off rapidly with the increase of $f$ at a fixed $z$. 
With the decrease of $f$, the outflow velocity $v_z$ increases rapidly due to 
strong radiation pressure which blows up the matter as shown in Fig. 1c. The 
disk gets possibly truncated due to strong outflow having 
both radiation and gas at a region around $r \sim 17$ for $f \sim 0.4$, 
and $r \sim 10$ for $f \sim 0.7$. 

In general, an increase of $f$ leads to
the inefficient cooling. This renders the disk to be puffed-up and more quasi-spherical. 
As a result, the disk angular momentum decreases due to its extraction by the outflow/jet (see Figs. 1b,c). 
However, at higher $f$ ($> 0.5$), the system becomes radiatively very inefficient,
which may result in the decrease of the possible outflux with an increase of $f$
rendering an increase of the flow angular momentum.
At low $\alpha$ ($\sim 0.01$) when the residence time of 
the infalling matter in the disk is high, angular 
momentum of the system is such that the disk becomes centrifugally 
dominated. At this stage, with the decrease of $f$, angular momentum 
may increase resulting in the radial and vertical velocity of the flow
to enhance significantly in order to overcome the strong centrifugal barrier.
The outflow is then centrifugally dominated. 

The Bernoulli's number in Fig. 3 
is similar to that of the velocity profiles in Fig. 1. 
$B_E$ is always positive and high at low 
$f$, which indicates the plausibility of the outflow to be very strong, and 
falls off rapidly with $r$. The probability of the outflow and jet is low 
at high $f$. With an increase of $z$, $B_E$ initially decreases. This 
is due to the fact that the first term of the potential $\phi_G$, which  
dominates at small $z$, is attractive in nature. Then $B_E$ 
gets a kick as the repulsive part of $\phi_G$ dominates with the increase in $z$.

\begin {figure}
\centering
\includegraphics[width=\textwidth,angle=270]{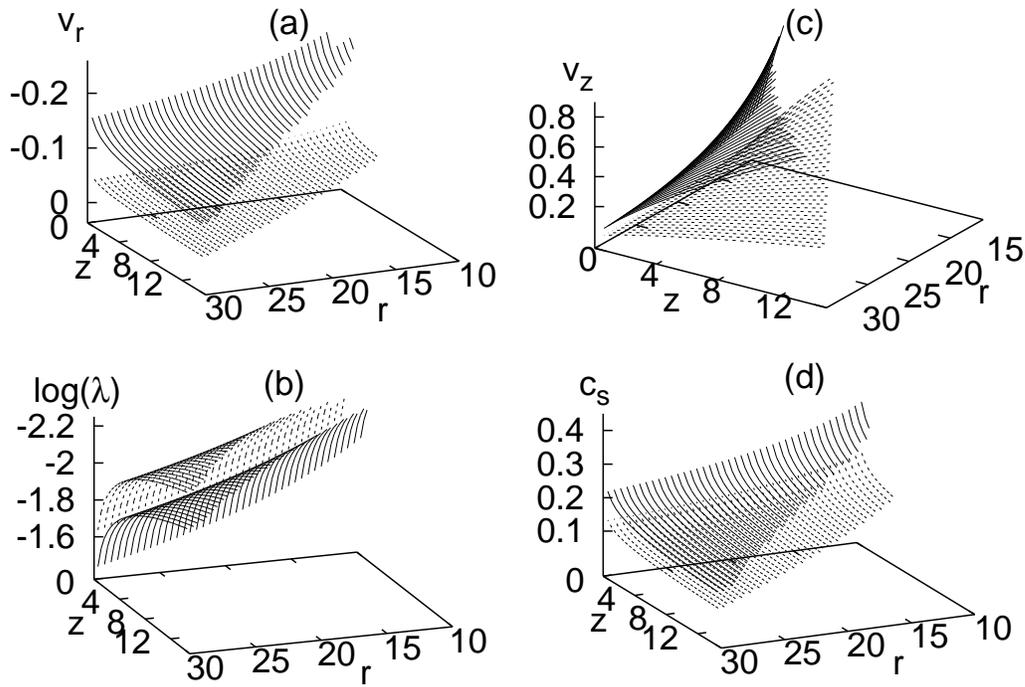}
\caption{
Variation of (a) velocity, (b) specific angular momentum, 
(c) vertical velocity, (d) sound speed, as functions of radial 
and vertical coordinates for super-Eddington accretion flows. 
Solid and dashed sheets are for 
$f = 0.4, 0.7$ respectively. Other parameters are $\alpha = 0.05$, $\gamma \sim 1.4$ and corresponding $\beta \sim 0.3$.
}
\label{Fig1}
\end{figure}

\subsection{Sub-critical accretion regime}

For highly sub-critical accretion flows, which are associated with very low density plasma, the possibility of transfer of viscous energy from ions to electrons due to the Coulomb collisions is very negligible. This results in a gas pressure dominated geometrically 
thick accretion disk. 
The flows have strong advection due to inefficient cooling and are optically thin 
(Narayan \& Yi \cite{ny94,ny95}). To analyse our result we choose $\gamma = 1.6$ 
which corresponds to $\beta \sim 0.89$ and $f = 0.9$.  
Figure 2 shows the profiles 
of the flow parameters as functions of $r$ and $z$ which are generally 
similar to those in the super-critical flows. 
The velocity profiles signify that the magnitudes of $v_r$ and $v_z$ 
are much less compared to that in the super-critical 
accretion flows. In low mass accretion flows, the disk may get 
truncated at much nearer to the central star.

\begin {figure}
\centering
\includegraphics[width=\textwidth,angle=270]{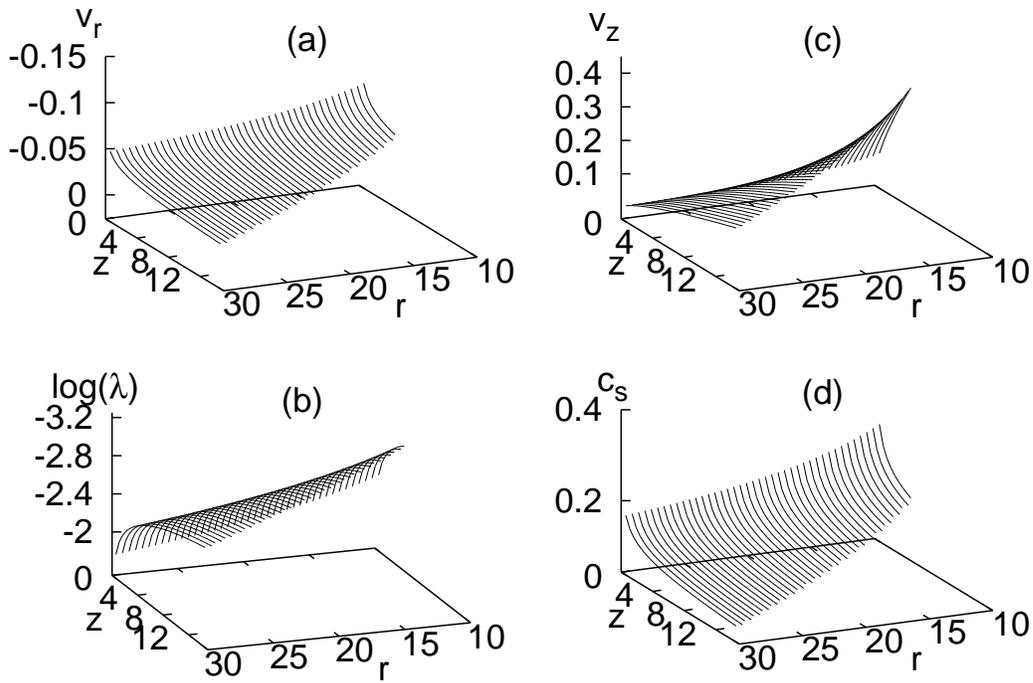}
\caption{
Same as in Fig. 1, but for sub-Eddington accretion flows.
Other parameters are $\alpha = 0.05$, $f = 0.9$, $\gamma \sim 1.6$ and corresponding $\beta \sim 0.89$.
}
\label{Fig2}
\end{figure}

\begin {figure}
\centering
\includegraphics[width=\textwidth,angle=270]{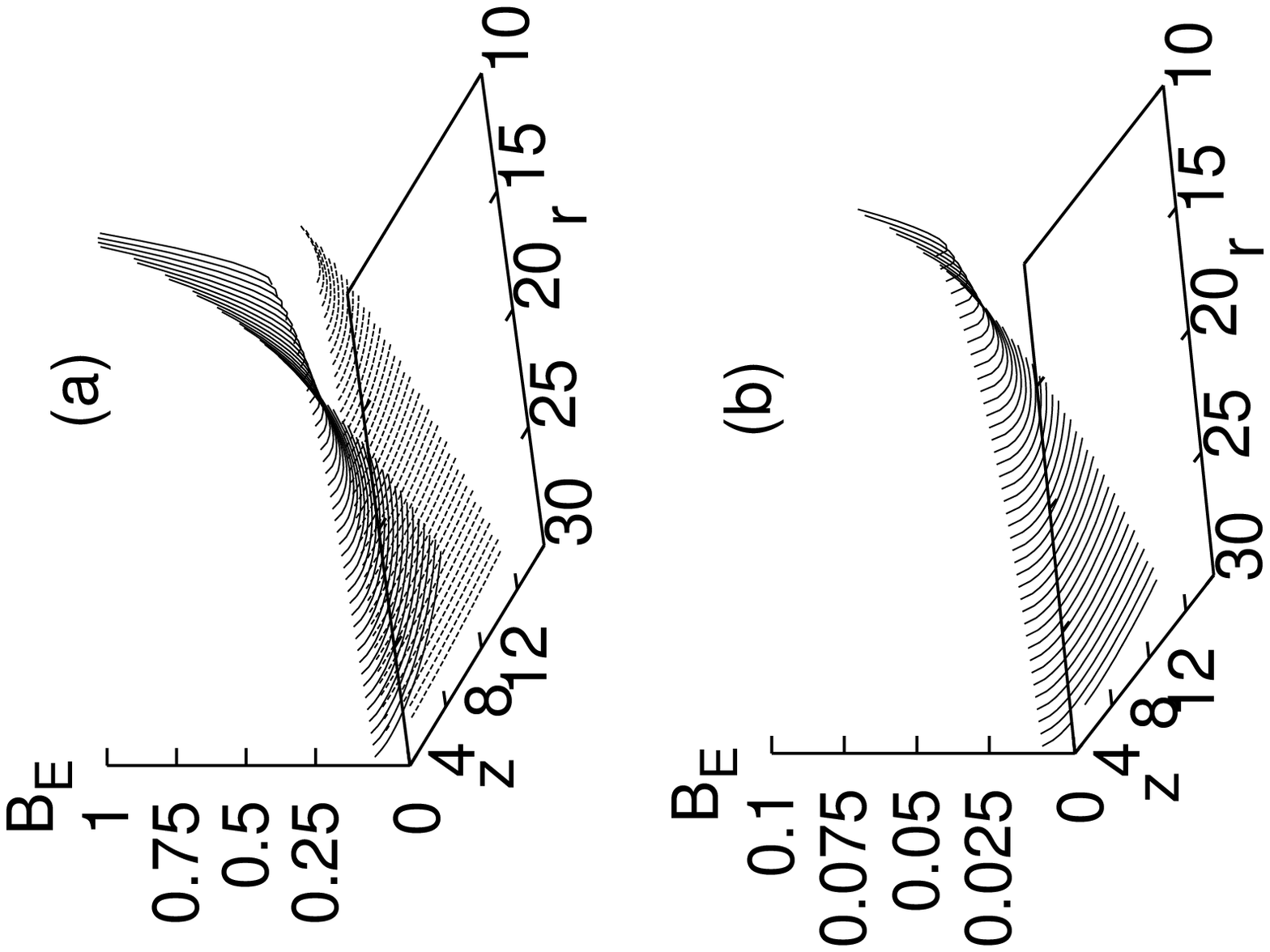}
\caption{
(a) Variation of Bernoulli's constant
for high mass accretion flows. Solid and dashed sheets are for $f = 0.4, 0.7$ respectively.
Other parameters are $\alpha = 0.05$, $\gamma \sim 1.4$ and corresponding $\beta \sim 0.3$.
(b) Same as in (a), but for low mass accretion flows. $\alpha = 0.05$, $f = 0.9$, $\gamma = 1.6$ and corresponding
$\beta \sim 0.89$.
}
\label{Fig3}
\end{figure}

\section{Summary}

We have presented a self-consistent model of accretion-induced 
outflow and then jet. We have established our model equations 
in a more general way, than done earlier, without making any hypothesis, 
and without restricting ourselves to the Keplerian geometry. Our equations uphold 
the conservation laws as the outflows and jets extract matter, energy and 
angular momentum from the inflowing matter. In its analytical self-similar form, it is more easy to
analyse and study the family of solutions (with variation of $\alpha \& f$) and to understand the significance of individual terms on the coupled dynamics of the 
flow. The only limitation 
we have kept is to ignore the importance of magnetic field in the
disk-outflow system. While not including magnetic
field is an assumption, the outflows and then jets in ULX are expected to 
emerge due to strong radiation pressure. Therefore, the collimation of jet in
ULX might not be magnetically linked (Jaroszy\'nski \& Abramowicz \cite{ja},
Fabrika \cite{fs}).
Therefore, for ULX and highly luminous AGN, the assumption of
neglecting magnetic effects could be quite appropriate.
We also do not aspire to describe the mechanism for formation
of jets, for that the inclusion of magnetic terms might be mandatory,
but try to understand the accretion flow dynamics with the
inclusion of the vertical flow. Moreover, to include magnetic field,
solve the equations, and obtain the solution in its present
form is beyond the scope. 

The new insights that we have provided in the work are:  \\
1) We have studied the complete set of axisymmetric Navier-Stokes equations 
for accretion-induced outflow analytically. \\
2) The vertical flow, which represents outflow, has been explicitly and self-consistently 
incorporated in our model, thus invoking a 2.5-dimensional accretion flow. \\
3) We do not assume hydrostatic equilibrium. \\
4) We have explicitly included $\phi z-$ and $rz-$ components of stress 
tensor apart from the usual $r \phi$-component in order to include
outflow dynamics into the disk. \\
5) All the flow parameters are considered to be functions of both 
$r$ and $z$ coordinates. We explicitly have shown, by order of magnitude 
analysis, which terms are relevant and which others can be discarded. 

Two extreme cases of the geometrically thick advective accretion disk 
consisting of super-critical and high sub-critical accretion flows have 
been studied. It shows that the dynamics of the system depends strongly on 
$f$. The model shows that the outflows and jets are less 
probable in sub-critical flows compared to that of super-critical flows. 
Although we have made a self-similar analytical study, it exhibits some  
reasonable features in understanding the dynamics of the accretion-induced outflow.

\acknowledgements
SG would like to thank the hospitalities of the
Department of Physics, Indian Institute of Science, where
most of the work has been done.
This work is partly supported by a project, Grant No. SR/S2HEP12/2007, funded 
by DST, India.

\label{lastpage}

\end{document}